# Mental Representations Constructed by Experts and Novices in Object-Oriented Program Comprehension

Jean-Marie Burkhardt*, Françoise Détienne*, and Susan Wiedenbeck **

\* Ergonomic Psychology Group, INRIA
Domaine de Voluceau, Rocquencourt, BP 105,
78153, Le Chesnay, cedex, France
Jean-Marie.Burkhardt@inria.fr, Francoise.Detienne@inria.fr

\*\* Computer Science and Engineering Department,
University of Nebraska
Lincoln, NE 68588-0115, USA
susan@cse.unl.edu

**ABSTRACT** Previous studies on program comprehension were carried out largely in the context of procedural languages. Our purpose is to develop and evaluate a cognitive model of object-oriented (OO) program understanding. Our model is based on the van Dijk and Kintsch's model of text understanding (1983). One key aspect of this theoretical approach is the distinction between two kinds of representation the reader might construct from a text: the textbase and the situation model. On the basis of results of an experiment we have conducted, we evaluate the cognitive validity of this distinction in OO program understanding. We examine how the construction of these two representations is differentially affected by the programmer's expertise and how they evolve differentially over time.

**KEY WORDS** mental representation, situation model, program model, object-oriented programming, program comprehension, text comprehension, expertise

## 1. OBJECTIVES

The object-oriented (OO) paradigm is growing fast in popularity, but not enough scientific evidence has been amassed about it. The research that exists is mostly focused on program design and reuse (see for example: Détienne, 1995; Pennington, Lee and Rehder, 1995). Furthermore, there is, as far as we know, no empirical work on the comprehension processes of OO programmers. Our objective is to investigate the mental representations constructed by programmers in the course of tasks involving the comprehension of software, e.g., reuse of parts of past designs and documentation of code. Previous studies on the comprehension of software texts were carried out in the context of procedural or functional languages. Based on the van Dijk and Kintsch's model of text understanding (1983), Pennington (1987a, 1987b) developed and tested a model of procedural program comprehension. One key aspect of the van Dijk and Kintsch's model is the distinction between two kinds of representation the reader might construct from a text: (1) the textbase which refers to what is said in the text and how it is said and (2) the situation model which represents the situation which is referred to by the text. The purpose of this work is to develop and evaluate a cognitive model of OO program understanding. More specifically, we will evaluate the cognitive validity of the distinction between the textbase (or program model) and the situation model in OO program understanding.

Section 2 presents our theoretical framework for studying OO program understanding and our research questions. In section 3, we present the methodology used in our empirical study. In section 4 we present and discuss the results of this study.

## 2. THEORETICAL FRAMEWORK

### 2.1. The mental model approach to text comprehension

Current models of text comprehension (Johnson-Laird, 1983; Schmalhofer and Glavanov, 1986; van Dijk and Kintsch, 1983) assume, in addition to the surface form (or verbatim) representation, two distinct, but interacting, levels of cognitive representation: (1) the textbase or propositional representation and, (2) the situation model or mental model. Level 1 is the linguistic representation of the text. It is isomorphic with the text structure and reflects what is contained in the text at a propositional level, i.e., it represents the microstructure and the macrostructure of the text. Level 2 is seen as an a-linguistic representation of the text that is isomorphic or homomorphic with the situation described by the text. Theory assumes that the propositional representation is built by mean of automatic processes from the verbatim representation. The building of the situation model, by contrast, is optional (Mills, Diehl, Birkmire, and Mou; 1995). It is produced by inferences and it makes extensive use of the subject's existing domain knowledge.

### 2.2. The mental model approach applied to procedural program comprehension

Pennington (1987a, 1987b) adapted van Dijk and Kintsch's text comprehension model to procedural program comprehension and tested it empirically. Pennington distinguished between two different mental representations which may be built while comprehending a program: 1) the **domain model** which is equivalent to van Dijk and Kintsch's situation model and reflects entities of the problem domain and their relationships, and 2) the **program model** which is equivalent to van Dijk and Kintsch's propositional textbase and reflects the text-based representation of the program.

Pennington argued that control flow and elementary operations information belong to the program model, while function and data flow information belong to the domain model. Her experimental paradigm was to give subjects a program to read for a limited time and then ask them questions reflecting different information categories, presumed to make up the situation and program models. The correctness of the responses to the questions then served as an indicator of the nature of the representation. Her experiments (Pennington, 1987a; 1987b) generally supported the dual model. She showed that control flow and elementary operations representations, which make up the program model, emerge first during program comprehension, perhaps because the programmer initially segments the elementary operations of one line or less into larger text structure units representing control flow. Function and data flow knowledge, which makes up the domain (or situation) model, emerge later with continued processing of the program.

### 2.3. Taking into account the OO nature of programs and the size of programs

There are several limitations to this approach related to the procedural nature of the languages used by Pennington and to the small size of the program she used. Pennington conducted her experiments with procedural languages and she did not examine at all representations about objects or even data structures. However, objects are central entities in OO programs and the construction of the representation of objects should be taken into account in a model of OO program understanding. We assume that the representation of objects is part of the situation model inasmuch as it reflects the objects of the problem situation.

Furthermore Pennington's model accounts for understanding of short programs but does not scale up easily to larger programs. Two aspects are not accounted for: the representation of delocalised plans and the representation of text macrostructure. Pennington assumes that the reader uses plan knowledge to construct the situation model. A plan is a set of actions that, when placed in a correct order, achieves some desired goal. Programmers have knowledge about patterns of program instructions which typically go together to accomplish certain functions or goals (Soloway, Ehrlich and Bonar, 1982). Pennington assumes that plan representations of a program are primarily based on data flow relations. In long programs, particularly in OO programs, it happens that many plans are delocalised. According to Rist (1996), plans and objects are orthogonal in OO systems. A plan can use many objects and an object can be used in many plans. In an OO system, the actions in a plan are encapsulated in a set of routines, and the routines are divided among a set of classes and connected by control flow. In our model we take the view that the construction of these complex delocalised plan representations is primarily based on client-server relationships, in which one object processes and supplies data needed by another object.

The macrostructure of long programs is not accounted for at all in Pennington's model, at the level of the program model. She accounts for the representation of elementary units, such as elementary operations and control flow between these operations, but she does not account for the representation of larger text units such as routines. Our model considers these text units as reflected by the macrostructure of the program model.

## 2.4. Proposed model of comprehension of object-oriented programs

As in Pennington's model, our **situation model** contains information about goals and data flow. In order to take into account the OO nature of programs and the size of programs, we have added information about objects as well as client-server relationship between objects. Information about objects and goals represents the static aspects of the problem solution, whereas information about data-flow and client-server relationships represents more dynamic aspects of the solution to the problem.

To summarize, the static aspects of the situation model refer to:
(1) the problem objects which directly model objects of the problem domain;
(2) the relationship between those objects, i.e., the inheritance/composition relationships between objects;
(3) the computing or reifed objects (e.g., a string class which is not a problem domain object). They are also represented at this level inasmuch as they are necessary to complete the representation of the relationships between problem objects, i.e., they bundles together program-level elements needed by the domain objects;
(4) the main goals of the problem. They correspond to functions of the program viewed at a high level of granularity. They do not correspond to single program units. The complex plan which realizes one goal is usually a delocalized plan in an OO programs.

The dynamic aspects of the situation model represent the communication between objects at a high level of granularity and the communication between variables at a fine level of granularity. These relationships trace the delocalized plans and the local plans involved in the problem solution. They are:
(1) Communications between objects correspond to client-server relationships in which one object processes and supplies data needed by another object. These connections between objects are the links connecting units of complex delocalized plans. In an OO system, the actions in a complex plan which performs a main goal are encapsulated in a set of routines, and the routines are divided among a set of classes and connected by control flow. Client-server relationships represent those connections.
(2) Communications between variables correspond to data flow relationships connecting units of local plans in a routine.

The **program model** contains two levels:
(1) at a microlevel, elementary operations constitute basic text units and control flow information constitutes the links between text units. Control flow, at this fine level of granularity, represents the control structure (either sequence, loop or test) linking individual operations;
(2) at a macrolevel, larger text units are represented. These are functions corresponding to units in the program structure, i.e., routines attached to objects.

As in text understanding theory, we assume that the construction of the situation model makes extensive use of the subject's existing domain knowledge. This representation is produced by inferences and is also a source of new inferences. Two kinds of knowledge, generic and episodic, and two knowledge domains, the problem domain and the programming domain, may be involved in this construction. Inferences may be drawn on the basis of schematic knowledge from the programming domain, such as plan knowledge. Knowledge about the problem situation, either generic knowledge or episodic knowledge, may also be activated and used as a source for inferences. On the contrary, the construction of the program model is based mostly on text structure knowledge and on local inferences for connecting propositions.

An important idea is that, whereas the construction of the textbase is systematic/automatic, the construction of the situation model is optional. Also, the building of the situation model requires time. It depends on the subject's knowledge and also on the task.

## 2.5. Research questions

We wish to evaluate the cognitive validity of the distinction between the program model and the situation model in OO program understanding. In particular, we wish to examine how the construction of these two kinds of representation is differentially affected by the programmer's expertise and how they evolve differentially over time.

Our first question is how expertise in programming affects the construction of the two representations. Here we must make clear that we do not manipulate expertise in the problem domain (all subjects have knowledge in this domain), but rather we manipulate the expertise in the programming domain. We compare expert programmers in OO programming with advance computer science students learning OO programming. According to our model, the expertise of subjects should affect the construction of the situation model but not the construction of the program model, provided that our novices are advanced students.

However, it is worth noting that a different hypothesis would be made by advocates of OOP. They have made strong claims about the naturalness, ease of use, and power of this design approach (Rosson and Alpert, 1990). It has been argued that there is a direct correspondence between the OO paradigm and the way people naturally think about problems (Borgida, Greenspan and Mylopoulos, 1986). If this is true, decomposition of a problem into objects may be easier in the OO paradigm because it is driven more by knowledge about world structure and less by knowledge about design schemas (or programming plans) representing classes of solutions in the programming domain. On this basis it could be argued that constructing the situation model, in particular the static part of it, would be easy in OO program comprehension, whatever the programmer's level of expertise. Thus, it would be expected that expertise in programming would not affect the construction of the situation model.

Our second research question concerns how the program model and situation model evolve over time. We examine the representations constructed after a first phase of comprehension then after a second phase of comprehension. The question is whether there is an order of construction of representations in OO program comprehension.

*Another research question is to examine the effect of the task (or purpose for reading) on the construction of the representations in OO program comprehension. We chose two tasks, the reuse task and the documentation task, because there are two realistics purpose for reading in program comprehension and because the involved problem solving component is more important in the former than in the latter of these tasks. This question will not be addressed directly in this paper as the data are still under process.*

In order to analyze these research questions we conducted an experiment on OO program understanding by experts and novices. Our experimental paradigm is similar to the one used by Pennington. Subjects answered questions after having studied a program for a certain period of time. The question categories were revised in accordance with our model.

## 3. DESIGN AND METHODOLOGY

### 3.1. Experimental design

A three-factor mixed design was used, as shown in Figure 1. The between subjects factors were expertise (OO expert vs. OO novice) and task orientation (documentation vs. reuse). The within subjects factor was phase (preliminary comprehension phase vs. task performance phase) and information category. Half the subjects were given a documentation task and half a reuse task.

The comparison between question set 1 after the preliminary comprehension phase and question set 2 after the task performance phase addressed the research question about the order of acquisition of knowledge in comprehension. Data came from both correctness of responses to questions and reaction times. The comparison between experts and novices provided data relevant to the research question about comprehension differences between expert and novice OO programmers.

### 3.2 Subjects

The subjects were 30 object-oriented experts and 21 object-oriented novices. The experts were professional programmers with experience in object-oriented design with C++. The novices were advance computer science students who were experienced in C but had only a basic knowledge of object-oriented programming and C++.

|  | OO experts | | OO novices | |
|---|---|---|---|---|
|  | Documentation task | Reuse task | Documentation task | Reuse task |
| Preliminary comprehension phase | Study pgm with documentation orientation, then question set 1 | Study pgm with reuse orientation, then question set 1 | Study pgm with documentation orientation, then question set 1 | Study pgm with reuse orientation, then question set 1 |
| Task performance phase | Documentation of pgm, then question set 2 | Reuse to solve target problem, then question set 2 | Documentation of pgm, then question set 2 | Reuse to solve target problem, then question set 2 |

Figure 1 Experimental design

Thirty of the subjects were speakers of English and 21 were speakers of French.

### 3.3 Materials

The materials consisted of a database program of approximately 550 lines which managed personnel, student, and course information for a small university. The program was composed of 10 classes. It was written in object-oriented C++ and presented in 23 files. The domain of the problem allowed us to write a program which took good advantage of the OO paradigm, including ease of conceptualization in terms of objects, classes, and inheritance. As in Pennington's study (1987a), little documentation was included in the text of the program. During the task performance phase, reuse oriented subjects were given a variation of the library problem (Wing, 1988) to design and implement. *This problem was partially isomorphic to the database program and allowed for reuse by template copying and modification or by inheritance.* Documentation oriented subjects were asked to comment the code for the use of another programmer who would later maintain it.

Two matched sets of yes/no questions were developed, one to be used at the end of each phase. The questions fit conceptually into two classes targeting knowledge making up the program model and the situation model. Three information categories reflected information composing the program model (see **Table 1**).

| Elementary operations |
|---|
| Does the program contain the code fragment: if ((number == search) || (name == search)) return TRUE; else return FALSE;? |
| **Control flow** |
| In "initialize" are the professors initialized before the courses ? |
| **Elementary functions** |
| Does "Collection::maintain" print out a list and ask the user to input a selection.? |

Table 1 Example of questions from categories composing the program model

| Problem objects |
|---|
| Does the program define a "Schedule" class? |
| **Computing objects** |
| Does the program define a "Collection" class? |
| **Object relationships** |
| Does the "Researcher" class inherit from the "Employee" class? |
| **Goals** |
| Does the program allow you to create a new schedule for an upcoming semester? |
| **Client-server** |
| Does the "Schedule" class call a member function of the "Course" class? |
| **Data flow** |
| In "Schedule::maintain" does the value of "selection" affect the value of "offerings"? |

Table 2 Example of questions from categories composing the situation model

Six information categories reflected information composing the situation model (see **Table 2**). The first four represent the static part of the situation model and the last two represent the dynamic part.

Two matched questionnaires were used. Each questionnaire contained 54 questions (3 yes questions * 3 no questions * 9 question categories). Two versions, a French and an English version, of each questionnaire were created. The order of presentation of the questionnaires was counterbalanced. The questions were randomized for presentation to subjects.

### 3.4 Procedure

Experts and novices were assigned randomly to the documentation or reuse groups. They were given an orientation to study the program for later reuse or documentation, as appropriate. Subjects were then given the database program and asked to study it for 35 minutes. Verbal protocols were collected. After this preliminary comprehension phase oriented by documentation or reuse goals, subjects answered the first question set on-line. Correctness and reaction times to the questions were recorded. In the performance phase, subjects were asked to carry out the documentation or reuse task for 90 minutes. Again verbal protocols were collected. Finally, subjects answered the second set of comprehension questions.

*Subjects were provided with a paper version as well as an electronical version of the program files. They were allowed to run the program but could modify it only in the second phase of the experiment.*

### 4. RESULTS

In this section we first present the global results with respect to the factors under study: expertise, phase, and information category. Then we present more detailed results concerning the model of comprehension of OO programs presented earlier. Most of the analyses involve the correctness of responses to the questions, but reference is made to reaction times when appropriate. Whatever the comparison we made, we calculated the scores considering 6 as the maximum correctness value. We do not report results from the protocol analysis.

As indicated in the methodology section, each subject was assigned to the French or English form of the questionnaires, as appropriate. A preliminary analysis showed few effects of language and those restricted largely to the reaction time data, so language is not treated further here. Subjects were randomly assigned to a reuse or a documentation task. Task is not analyzed here; for the current purposes the important point is that both tasks were strongly comprehension-dependent.

Assignment of subjects to novice or expert levels was initially done on the basis of a self-report questionnaire. To verify the group assignments a posteriori, we carried out a preliminary analysis. A qualitative data analysis was made on subjects responses patterns by using a dynamic cluster method. The results strongly supported our original grouping.

Initially a three-way mixed model Analysis of Variance was performed on the number of correct responses to questions. The between subjects factor was expertise (novice or expert). The within subjects factors were phase (1=preliminary or 2=task performance) and question category (nine categories as described in section 3.3). The results showed that there was a significant overall effect of expertise ($m_{expert}$ = 4.23, sd = 1.32; $m_{novice}$ =3.89, sd = 1.29; $F(1, 44)$ = 11.08, p < .0018). There was a significant effect of phase ($m_{phase1}$ = 3.93, sd = 1.34; $m_{phase2}$ = 4.25, sd = 1.28; $F(1, 44)$ = 23.87, p < .0001). Category was significant, as well ($F(8, 352)$ = 36.68, p < .0001). The two-way interaction of expertise and category was significant ($F(8, 352)$ = 3.16, p < .0018). The two-way interaction of expertise and phase was not significant, nor was the two-way interaction of phase and category, nor the three-way interaction. See Figure 2 for the means of the interactions.

Our model of OO program comprehension contrasts the situation model with the program model. We carried out further analyses to test this model. The model-based analysis compared the six information categories making up the situation model (problem objects, computing objects, object relationships, goals, client-server relationships, and data flow) with the three information categories making up the progam model (elementary operations, control flow, and elementary functions). We found an overall difference between the situation and progam models ($m_{situation}$ = 4.36, sd = .61; $m_{program}$ = 3.55, sd = .65; $F(1, 43)$ = 110.94, p < .0001). Scores for the situation model improved significantly between phase 1 and phase 2 ($m_{situation-p1}$ = 4.17, sd = .65; $m_{situation-p2}$ = 4.55, sd = .51; F = 10.83, p < .002), while scores for the program model did not. Experts had significantly higher scores on the situation model than did novices ($m_{situation-expert}$ = 4.56, sd = .54;

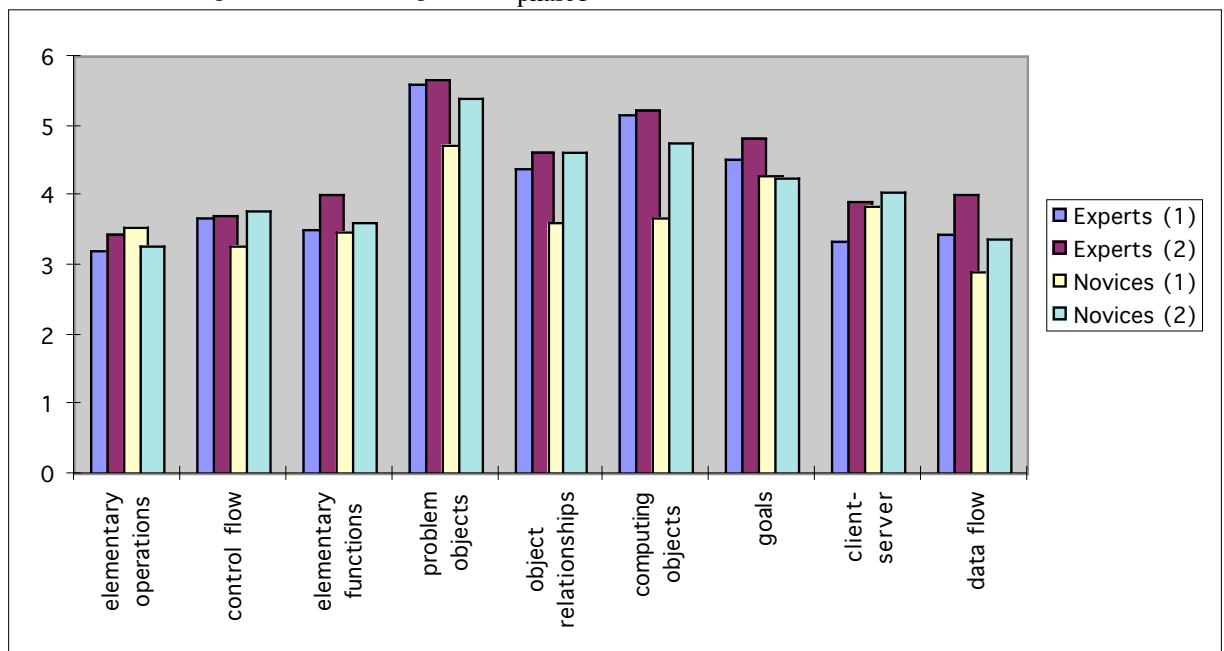

Figure 2 Mean correctness score by question category for novices and experts in phase 1 and in phase 2 ( max. = 6)

$m_{situation-novice}$ = 4.10, sd = .61; F(1, 44) = 19.17, p < .0001). However, there was no expert/novice difference on the program model.

As described above, the situation model may be considered to have two aspects: static information vs. dynamic information. To investigate the representation further we compared these two with each other and with the program model. The static situation knowledge contained problem objects, computing objects, object relationships, and goals. The dynamic situation knowledge contained client-server relationships and data flow. There was a significant overall difference between the static and dynamic information ($m_{static}$ = 4.74, sd = .68; $m_{dynamic}$ = 3.62, sd = .96; F(1, 45) = 106.20, p < .0001). Also, the reaction times (corrected for question length) for the program objects and goals categories of the static part were faster than any other categories. There was a significant difference between phase 1 and phase 2 for static information ($m_{static-p1}$ = 4.57, sd = .71; $m_{static-p2}$ = 4.90, sd = .59; F(1, 45) = 7.15, p < .0104) and also for dynamic information ($m_{dynamic-p1}$ = 3.38, sd = 1.09; $m_{dynamic-p2}$ = 3.85, sd = .75; F(1, 45) = 9.36, p < .0037). For the static information experts scored significantly higher than novices ($m_{static-expert}$ = 5.00, sd = .49; $m_{static-novice}$ = 4.36, sd = .73; F(1, 46) = 34.23, p < .0001). However, for the dynamic information there was no difference between experts and novices. There was a significant overall difference between the static part of the situation model and the program model (means given previously; F(1,45) = 237.92, p < .0001). Finally, for the comparison of the dynamic part of the situation model and the program model, there was no significant difference.

## 5. DISCUSSION

Our results tend to show that the distinction between the situation model and the program model has cognitive validity for OO programmers. First, our findings show that there is an overall difference between these two representations. The situation model appears to be stronger (based on the correctness of responses) than the program model. Second, our findings show that these two representations are affected differentially by the expertise and phase factors. Expertise has an effect on the construction of the situation model, whereas it has no effect on the construction of the program model. Furthermore, phase affects the construction of the situation model but not the program model, i.e., there is an enrichment of the situation model over time whereas the program model remains "stationary".

Our results also indicate an overall difference between the static and dynamic parts of the situation model. Based on the correctness scores, the static part of the model appears to be better developed than the dynamic part in the mental representation of OO programmers. The expertise factor affects the construction of only the static part, whereas the phase factor affects the construction of both the static and dynamic parts. In particular, experts develop a better static situation model than do novices. Both static and dynamic parts improve over time. Interestingly, the dynamic part of the situation model is not better developed than the program model. This implies that the marked difference between the situation model and the program model is accounted for by the effect of the static part.

According to our results, the situation model is more fully developed than the program model, even in an early phase of comprehension. This contrasts with the results of Pennington (1987a, 1987b) for procedural programmers. She showed that the program model developed more fully in the early stages of comprehension, whereas the situation model emerged later, after performance of a meaningful programming task. Perhaps this difference between our results and Pennington's can be explained by the programming paradigm. It appears that the OO paradigm, with its emphasis on objects and relationships of objects, may facilitate the construction of a situation model earlier in program comprehension. This is consistent with the naturalness claims of advocates of the OO paradigm. On the other hand, the situation model constructed by experts is stronger than that constructed by novices. This suggests that the construction of the situation model is not based solely on problem domain knowledge but also on knowledge acquired through experience in the programming domain. These latter findings are consistent with results on object-oriented design (Détienne, to appear) which show a greater benefit of this paradigm for expert designers than for novice designers.

With respect to the program model, our results show that it is not enriched over time as is the situation model. Rather it remains essentially constant. Perhaps this ceiling effect is related to the size of the program. In a larger program, it may be difficult to keep in working memory a representation of program level information. This resource limitation hypothesis requires further verification. Another possible explanation of lack of growth of the program model is the comprehension strategy of the programmer. A programmer may concentrate on certain parts of a program at the expense of others, as observed in the "as-needed" comprehension strategy (Koenemann and Robertson, 1991; Littman,

Pinto, Letovsky and Soloway, 1986). The use of an as-needed strategy appears to be task related. In further analyses, we will use our quantitative and qualitative data together to understand the relation between the construction of the mental representation and comprehension strategies.

In our results, both the static and the dynamic parts of the situation model evolved, becoming more elaborated over time. However, the static part always remained more developed than the dynamic part, even for experts. The program objects and goals were also the more accessible, as shown by reaction times. The dynamic part reflects the mental representation about plans, either local or delocalized. It appears that the OO paradigm facilitates the construction of the situation model most strongly in its static part. It may be hypothesized that the dynamic part is important for linking the static aspects of the situation model to the program model. Our results suggest that the OO paradigm does not strongly support this linking of the dual mental representations, at least not early in program comprehension.

## 6. LIMITATIONS AND FUTURE DIRECTIONS

There are several limitations of this study related to the methodology. Subjects worked with a single program which implemented a database. To generalize the results it is necessary to repeat the study with other programs *in other problem domains*. Furthermore, while the program was larger than often used in this kind of study, it was still a small program by industrial standards. Thus, we do not know whether the mental representation of a much larger program would conform precisely to what we found here. In our study subjects worked with the program for approximately 2 hours, and most did not have time to finish the reuse or documentation task they were given. We might have observed further evolution of the mental representation if they had worked with the program over a longer time.

One future direction will be to analyze the influence of our task factor on program comprehension. As shown in recent studies on text comprehension (Mills et al. 1995), the purpose for reading exerts an important influence on the construction of the mental representation. We expect that similar effects will be found in the domain of programming. In addition to analyzing the effects of a reuse or documentation task, we are interested in extending this study to other comprehension-related tasks, such as program modification.

Another direction of this research will be to analyze the comprehension strategies used by subjects performing different tasks and subjects at different levels of expertise. We plan to use qualitative data from our think-aloud protocols to link specific comprehension strategies to outcomes in the development of the mental representation.

## REFERENCES

Borgida, A., Greenspan, S., and Mylopoulos, J. (1986). Knowledge representation as the basis for requirements specifications. In C. Rich and C. R. Waters (Eds): *Readings in Artificial Intelligence and Software Engineering*. Los Altos, CA: Kaufmann. 561-570.

Détienne, F. (1995) Design strategies and knowledge in object-oriented programming: effects of experience. *Human-Computer Interaction*, 10 (2 & 3), 129-170.

Détienne, F. (to appear) Assessing the cognitive consequences of the object-oriented approach: a survey of empirical research on object-oriented design by individuals and teams. Interacting with Computers.

Johnson-Laird, P. N. (1983). *Mental models: Towards Cognitive Science of Language, Inference, and Consciousness*. Cambridge: Cambridge University Press.

Koenemann, J. and Robertson, S.P. (1991). Expert problem solving strategies for program comprehension. In S. P. Robertson, G. M. Olson, and J. S. Olson (Eds): *CHI'91 Conference Proceedings*, NY: ACM. 125-130.

Littman, D. C., Pinto, J., Letovsky, S., and Soloway, E. (1986). Mental Models and Software Maintenance. In E. Soloway and S. Iyengar (Eds): *Empirical Studies of Programmers*, Norwood, NJ: Ablex. 80-98.

Mills, B. C., Diehl, V. A., Birkmire, D. P. & Mou, L-C. (1995) Reading procedural texts: effects of purpose for reading and predictions of reading comprehension models. *Discourse Processes*, 20, 79-107.

Pennington, N. (1987a). Comprehension strategies in programming. In G.M. Olson, S. Sheppard, and E. Soloway (Eds): *Empirical Studies of Programmers: Second Workshop*. Norwood, NJ: Ablex. 100-113.

Pennington, N. (1987b). Stimulus Structures and Mental Representations in Expert Comprehension of Computer Programs. *Cognitive Psychology*, 19, 295-341.

Pennington, N., Lee, A. Y., & Rehder, B. (1995). Cognitive activities and levels of abstraction in procedural and object-oriented design. *Human-Computer Interaction*, 10 (2 & 3), 171-226.

Rist, R. (1996) System structure and design. In W. D. Gray, & D. A. Boehm-Davis (Eds): *Empirical Studies of Programmers, Sixth*. Norwood, NJ: Ablex. 163-194.

Rosson, M. B., and Alpert, S. R. (1990). The cognitive consequences of object-oriented design. *Human-Computer Interaction,* 5(4), 345-379.

Schmalhofer, F., and Glavanov, D. (1986). Three Components of Understanding a Programmer's Manual : Verbatim, Propositional, and Situational Representations. *Journal of Memory and Language,* 25, 295-313.

Soloway, E., Ehrlich, K., & Bonar, J. (1982) Tapping into tacit programming knowledge. *IEEE Transactions on Software Engineering*, SE-10, 595-609.

van Dijk, T.A. and Kintsch, W. (1983) *Strategies of Discourse Comprehension.* New York: Academic.

Wing, J.M. (1988) A study of 12 specifications of the library problem. *IEEE Software*, 5, 66-72.